\documentclass[
superscriptaddress,
preprint,
showpacs,preprintnumbers,
amsmath,amssymb,
aps,
floatfix,
]{revtex4-1}

\usepackage{graphicx}
\usepackage{dcolumn}
\usepackage{bm}
\usepackage[mathlines]{lineno}

\begin{document}

\title{Airy-soliton interactions in self-defocusing media with PT potentials}
\author{Zhiwei Shi}
\thanks{Corresponding author: szwstar@gdut.edu.cn}
\affiliation{School of Electro-mechanical Engineering, Guangdong University of Technology, Guangzhou 510006,P.R.China}
\affiliation{Department of Physics and Astronomy, San Francisco State University, San Francisco, California 94132, USA}
\author{Huagang Li}
\affiliation{Department of Physics, Guangdong University of Education , Guangzhou 510303, P. R. China}
\author{Xing Zhu}
\affiliation{Department of Physics, Guangdong University of Education , Guangzhou 510303, P. R. China}
\author{Zhigang Chen}
\affiliation{Department of Physics and Astronomy, San Francisco State University, San Francisco, California 94132, USA}
\affiliation{The MOE Key Laboratory of Weak-Light Nonlinear Photonics, TEDA Applied Physics Institute and Physics of School, Nankai University, Tianjin 300457, P. R. China}

\begin{abstract}
We investigate Airy-soliton interactions in self-defocusing media with PT potentials in one transverse dimension. We discuss different potentials in which the interacting beams with different phases are launched into the media at different separation distances. During interactions, there exist a primary collision region and a relaxation region accompanied by continuous interaction with the dispersed Airy tail. In the relaxation region, the beams exist soliton-like and breathers-like propagation. The beam width and mean power are influenced by initial separation, phase shift and modulation depth of PT potentials. Especially, the collision distance decreases with the spatial beam separation and the mean power possesses sinusoidal dependence on the phase shift.
\end{abstract}

\maketitle
\section{Introduction}

Over the years, there has been an increasing interest in the study that the diffraction can be eliminated through effective light-field regulation in many practical applications. Generally, there are two kinds of methods which can be used to offset the diffraction. One is to rely on the nonlinear (NL) effect in a medium. The
formation of spatial solitons can be thought as one of the most fundamental effect which gives rise to localized structures that propagate unchanged, stabilized by the balance between the diffraction and the NL effect~\cite{1}.
The other is non-diffracting beams in the free space. For instance, Airy optical beams, which exhibit self-accelerating, non-diffracting, and self-healing properties during propagation, were investigated theoretically and experimentally by Siviloglou \textsl{et al}~\cite{2,3} for the first time in 2007. These novel
properties of the Airy beams are ideally suited for various applications ranging
from particle micromanipulation~\cite{3-1,3-2}, self-bending plasma channels~\cite{3-3}, light bullets~\cite{3-4}, optical interconnects~\cite{3-5}, image signal transmission~\cite{3-6},
super-resolution imaging~\cite{3-7}, intrapulse Raman scattering~\cite{3-8}, to name just a few.

The potential arising from combining both methods to investigate
interactions of accelerating beams and solitons opens completely new aspects of research.
In the past decade, Airy beams in different NL media were widely studied, such as Kerr NL dielectrics~\cite{11,12,13,14,15}, photorefractive media~\cite{17}, nonlocal NL media~\cite{18,19}, and quadratic media~\cite{11,22}. Because of the existence of nonlinearities, self-trapped beams can be realized with Airy-like beams/pulses~\cite{24,25,26,27,28} and self-accelerating solitary-like waves can also be found~\cite{12,13,17,22,28-1,28-2,28-3}. Refs.~\cite{11,12,13,14,15,17,18,19,22,24,25,26,27,28,28-1,28-2,28-3} discussed non-diffracting beams all in uniform media. By considering modulated refractive index potentials (i.e., the media are no longer uniform) a new degree of freedom is added to the system which brings about exciting new effects of the Airy beam propagation, which has already been considered in a few theoretical and experimental studies. Efremidis studied the propagation of Airy beams in transversely linear index potentials with a gradient that is dynamically changing along the propagation direction~\cite{28-4} and the particular case of Airy-type optical waves that are reflected and transmitted by two generic classes of potentials ~\cite{28-5}. In inhomogeneous media with a linear gradient index distribution, Moya-Cessa \textsl{et al.} demonstrated that an Airy beam propagates in a straight line due to the balance between the inhomogeneous medium and the diffraction of the beam, in a similar way as a solitary wave in a NL inhomogeneous medium~\cite{28-6}. Makris \textsl{et al.} studied non-diffracting accelerating paraxial optical beams in periodic potentials, in both the linear and the NL domains~\cite{29}. Hu \textsl{et al} studied the behavior of Airy beams propagating from a NL medium to a linear medium~\cite{30} and in optically induced refractive-index potentials~\cite{31}.
The propagation dynamics and beam acceleration can also be controlled with positive and negative defects, and appropriate refractive index change~\cite{32}. Moreover, Dragana M Jovi\'{c} \textsl{et al} analyzed the influence of an optically induced photonic lattice on the acceleration of Airy beams~\cite{33,34}. However, the propagation dynamics of Airy beams in parity-time (PT)-symmetric potentials has thus far not been reported to our knowledge.

In optics, PT-symmetric potentials can be designed by introducing a
complex refractive-index distribution $n(x)=n_R(x)+in_I(x)$ ,where $n_R(x)=n_R(-x)$, $n_I(x)=-n_I(-x)$, and $x$ is the normalized transverse coordinate~\cite{35,36}. When NL is introduced, a novel class of NL self-trapped modes was found, and the interplay between the Kerr nonlinearity and the PT threshold was analyzed by Musslimani and co-worker~\cite{35}. Christodoulides group presented closed form solutions to a certain class of one- and two-dimensional NL Schr\"{o}dinger equations involving potentials with broken and unbroken PT symmetry~\cite{37}. We also showed that defocusing Kerr media with PT-symmetric potentials can support one- and two-dimensional bright spatial solitons~\cite{38}.

Moreover, the interactions between Airy pulse and temporal solitons at the same
center wavelength~\cite{42} or at a different wavelength~\cite{43} has been studied.
The interaction of an accelerating Airy beam and a solitary wave has also been investigated for integrable and non-integrable equations governing NL optical propagation~\cite{44}. Up to now, the interactions between an Airy beam and stable spatial beams in self-defocusing media with PT lattices have not been mentioned. When the two beams are placed in proximity to each other, with the Airy acceleration direction towards the soliton, interesting questions arise. For instance, will the soliton behave as an impenetrable barrier? Can the Airy probe control the soliton propagation
parameters? How do the corresponding parameters affect the interaction of the beams? These questions are addressed in this paper. In this paper, we address the above questions. Specifically, we investigate the dynamics of two one-dimensional interacting beams along the propagation direction. We discuss the influence of different physical parameters on the beam interaction, including the initial spatial beam separation, phase difference, and amplitude ratio between the beams, the modulation depth and the width of PT potentials.
The organization of the paper is as follows. We briefly introduce the theoretical model in Sec.~\ref{section:two}; in Sec.~\ref{section:three} and Sec.~\ref{section:four}, we discuss numerically Airy beam propagation and Airy-soliton interactions in PT potentials in details, respectively. Section~\ref{section:five} concludes the paper.
\begin{figure}[htb]
\centering
\fbox{\includegraphics[width=\linewidth]{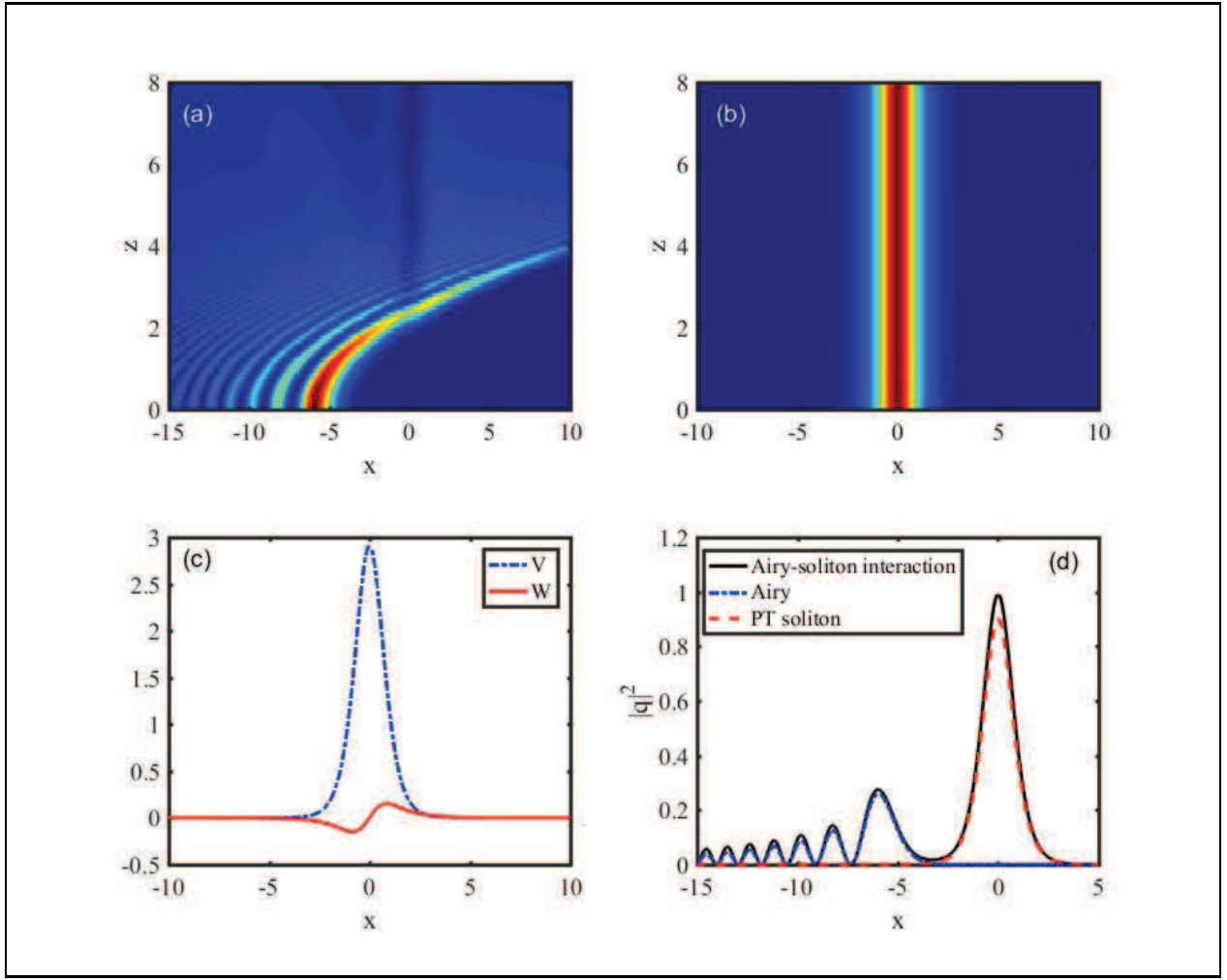}}
\caption{(color online) Intensity plots for the propagation of (a) a weak truncated Airy beam and (b) a normalized PT soliton. (c) The real $V(x)$ and imaginary $W(x)$ parts of PT potentials. (d) Exemplary initial launch conditions composed of both the Airy beam and the normalized soliton. The physical parameters are $\alpha=0.03$, $q_{s0}=1$, $q_{A0}=1$, $D=-5$, and $\theta=0$.}
\label{fig:one}
\end{figure}
\section{The theoretical model}
\label{section:two}
In a Kerr self-defocusing medium with PT-symmetric
potentials, the scaled equation for the propagation of a slowly-varying envelope $q$ of the optical electric field in one transverse dimension in the paraxial approximation is normalized NL Schr\"{o}dinger equation (NLSE)~\cite{37,38},
\begin{equation}
i\frac{\partial q(x,z)}{\partial z}+\frac{1}{2}\frac{\partial^2q(x,z)}{\partial x^2}+R(x)q(x,z)+\gamma|q(x,z)|^2q(x,z)=0,
\label{eq:one}
\end{equation}
where $z$ is the propagation distance, $R=V(x)+iW(x)$, and
$V(x)=V_0\sec\textrm{h}^2(x/d)$ and $W(x)=W_0\sec\textrm{h}(x/d)\tanh(x/d)$ are the real and imaginary components of the complex PT-symmetric potential, respectively. $V_0$ and $W_0$ are the amplitudes of the real and imaginary parts. $d$ denotes the width of PT potentials. Before proceeding to the interaction discuss, we review first the fundamental PT soliton and Airy beam properties.

\subsection{Spatial Airy beam definition}
Obviously, when $R=0$ and $\gamma=0$, Equation~(\ref{eq:one}) is simplified to the linear SE $i\partial q/\partial z+1/2\partial^2q/\partial x^2=0$. Here, one of the accelerating solutions of Equation (\ref{eq:one}) is the well-known Airy function with
the characteristic infinite oscillatory tail~\cite{2},
\begin{equation}
q(x,z)=Ai(x-\frac{z^2}{4})\exp[\frac{i}{12}(6xz-z^3)],
\label{eq:two}
\end{equation}
From this solution, it is easy to see that the trajectory is determined by the transverse
accelerating term $x-z^2/4$, so the beam propagates along a parabolic curve. The intensity of
an ideal Airy wave packet remains invariant during propagation, but the ideal Airy beam does not exist in reality for its infinite energy. To make it finite-energy, an input Airy beam
defined by $q(x,0)=\textrm{Ai}(x)\exp(\alpha x)$ evolves as~\cite{2,42}:
\begin{equation}
q(x,z)=Ai(x-\frac{z^2}{4}+i\alpha z)\exp[\frac{i}{12}(6\alpha z^2-12i\alpha z+6i\alpha z^2+6xz-z^3)],
\label{eq:three}
\end{equation}
where $\alpha\geq0$ is an arbitrary real decay parameter. Figure \ref{fig:one}(a)
shows an intensity plot for the propagation of a truncated Airy beam with $\alpha=0.03$.

\subsection{Spatial PT soliton definition}
If we assume $\gamma=-1$, the NLSE supports a PT soliton solution which can be described as~\cite{38},
\begin{equation}
q(x,z)=q_{s0}\sec\textrm{h}(x)\exp(i\rho\arctan(\sinh(x)))\exp(i\beta z),
\label{eq:four}
\end{equation}
where $\rho=W_0/3$, $q_{s0}=\sqrt{V_0-w_0^2/9-2}$, and the propagation constant of the soliton $\beta=1$. Figures \ref{fig:one}(b) and \ref{fig:one}(c)
show an intensity plot for the propagation of a PT soliton and the real $V(x)$ and imaginary $W(x)$ parts of PT potentials, respectively.

\subsection{Airy-soliton interactions}
To investigate the interaction of an Airy beam and a PT soliton, we take the initial beam as a superposition of two beams
\begin{equation}
q(x,0)=q_{s0}\sec\textrm{h}(x)\exp(i\rho\arctan(\sinh(x)))+q_{A0}\textrm{Ai}(x-D)\exp(\alpha (x-D))\exp(i\theta),
\label{eq:five}
\end{equation}
where $q_{A0}$ are the amplitude of the Airy beam. $D$ is the initial Airy beam position with respect to the soliton (launched at $z=0$), and $\theta$ controls the phase shift. When we vary these parameters, the soliton propagation must be affected. We demonstrate these interactions through numerical simulations using the Split Step Fourier Method. In our simulations, we choose a small truncation coefficient $\alpha=0.03$, which guarantees that all the Airy beams have the same energy for a given $q_{A0}$ value and only a small variation in peak
intensity at the point of collision~\cite{42}. Figure \ref{fig:one}(d) illustrates exemplary initial launch conditions composed of both the Airy beam and the normalized PT soliton, where we can say that the soliton overlaps or fuses with the Airy beam.

\section{Numerical results of Airy beams in PT potentials}
\label{section:three}
\begin{figure}[htb]
\centering
\fbox{\includegraphics[width=\linewidth]{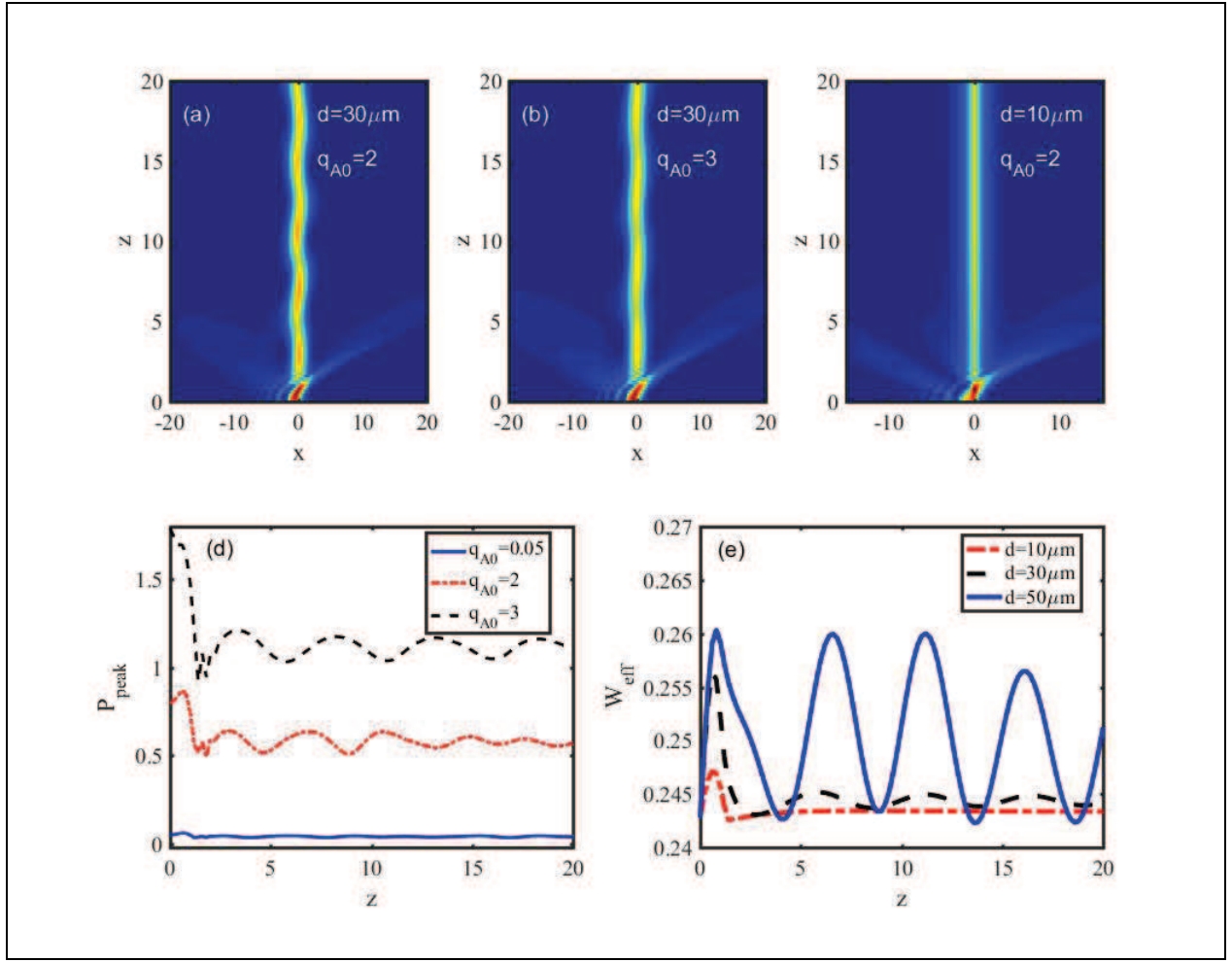}}
\caption{(color online) Intensity plots for the propagation of an Airy beam at (a) $q_{A0}=0.5$, $d=30\mu$m, (b) $q_{A0}=2$, $d=30\mu$m and (c) $q_{A0}=2$, $d=10\mu$m. $P_{peak}$ and $W_{eff}$ as a function of the propagation distance $z$ for different $q_{A0}$ (d) and $d$ (e) with $V_0=3$ and $W_0=0.3$.}
\label{fig:two}
\end{figure}
\begin{figure}[htb]
\centering
\fbox{\includegraphics[width=\linewidth]{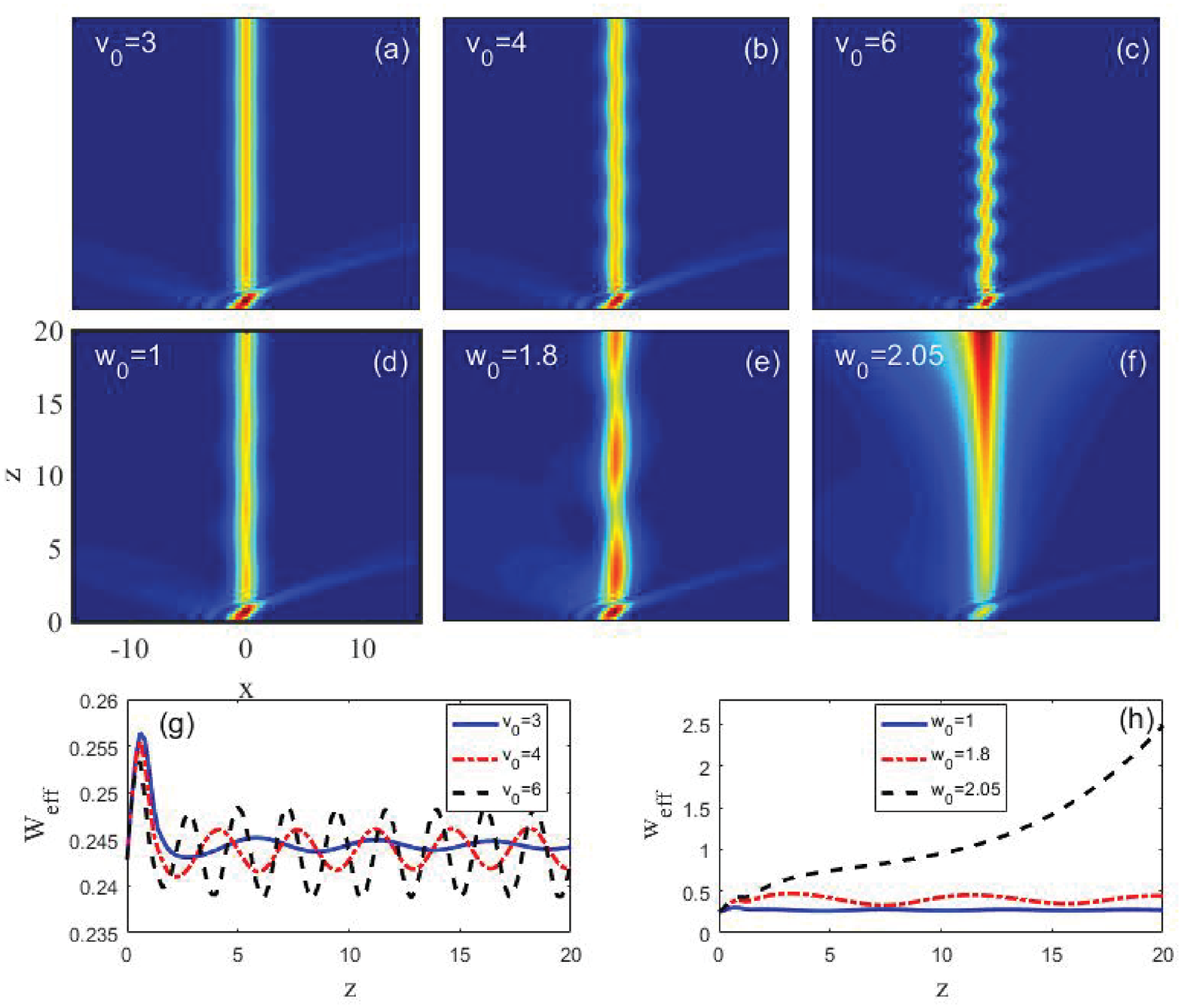}}
\caption{(color online) Intensity plots for the propagation of a Airy beam at (a) $V_{0}=3$, (b) $V_{0}=4$, and (c) $V_{0}=6$ with $W_0=0.3$ and (d) $W_{0}=1$, (e) $W_{0}=1.8$, and (f) $W_{0}=2.05$ with $V_0=3$. (g) The effective beam width $W_{eff}$ via the propagation distance $z$ for $W_0=0.3$. (h) $W_{eff}$ via $z$ with $V_0=3$. The other parameters are $q_{A0}=2$ and $d=30\mu$m.}
\label{fig:three}
\end{figure}
Before discussing Airy-soliton interactions, we only consider the propagation of Airy beams in a defocusing nonlinear medium ($\gamma=-1)$) with PT potentials. So, the input beam is $q(x,0)=q_{A0}\textrm{Ai}(x)\exp(\alpha x)$. We can see that Airy beams are divided into three parts due to the existence of PT potentials from Figs.~\ref{fig:two} and \ref{fig:three}. One part may be soliton-like with small oscillations; The other part stays the same self-accelerating as the input Airy beam; The last part may be reflected waves. Here, we only pay attention to the first part. The influence of the amplitude of Airy beams and the width of PT potentials on their propagation properties is shown in Fig.~\ref{fig:two}.
The amplitude of the beam oscillation decreases with the increasing $q_{A0}$, which can be seen from Fig.~\ref{fig:two}(a) and (b). This can be illustrated by using Newtonian mechanics.  $q_{A0}$  is equivalent to the beam mass. When an Airy beam interacts with a PT potential, the amplitude of the beam oscillation must decrease with the increasing $q_{A0}$ for the same acting force of PT potential. However, not only the peak intensity of the beam $P_{peak}$ but also the oscillation amplitude of $P_{peak}$ increase with $q_{A0}$, to see Fig.~\ref{fig:two}(d). In addition, we also see that the oscillation amplitude of $P_{peak}$ gradually becomes smaller during propagation because of the existence of beam decay. One can also find the effect of the width of PT potentials $d$ on Airy beams. When $d$ is smaller ($d=10\mu$m), the soliton may form when $z>8$ (Fig.~\ref{fig:two}(c)). While $d$ increases ($d=30\mu$m), the beam oscillations (the soliton-like dynamics) shown in Fig.~\ref{fig:two}(b) may appear. Fig.~\ref{fig:two} (e) shows that the effective beam width $W_{eff}=\int_{-\infty}^{+\infty}|q|^2/e^2$ varies with $z$ for different $d$, where $e$ is $e$ exponential. We can find that $W_{eff}$ is almost constant for $d=10\mu$m, but it is oscillating when $d=30\mu$m and $d=50\mu$m. Moreover, the amplitude of oscillation and $W_{eff}$ increase with the width of PT potentials $d$. That is to say, as $d$ is smaller, the Airy beam would be much more tightly localized because the binding force of PT potentials becomes stronger. Fig.~\ref{fig:three} illustrates how the amplitudes of the real and imaginary parts of PT potentials affect the beam propagation properties. Firstly, One can say that the amplitude of oscillations of soliton-like largens and the oscillation period shortens from Figs.~\ref{fig:three}(a)-(c) and Fig.~\ref{fig:three}(g) when $V_0$ increases. Secondly, when we change $W_0$, the properties of the beams are interesting. At $W_0=1$, the soliton-like may appear in Fig.~\ref{fig:three}(d); At $W_0=1.8$, the breathers-like may take shape in Fig.~\ref{fig:three}(e); At $W_0=2.05$, the beam may diffract in Fig.~\ref{fig:three}(f).
These results can be verified in Fig.~\ref{fig:three}(h).
\begin{figure}[htb]
\centering
\fbox{\includegraphics[width=\linewidth]{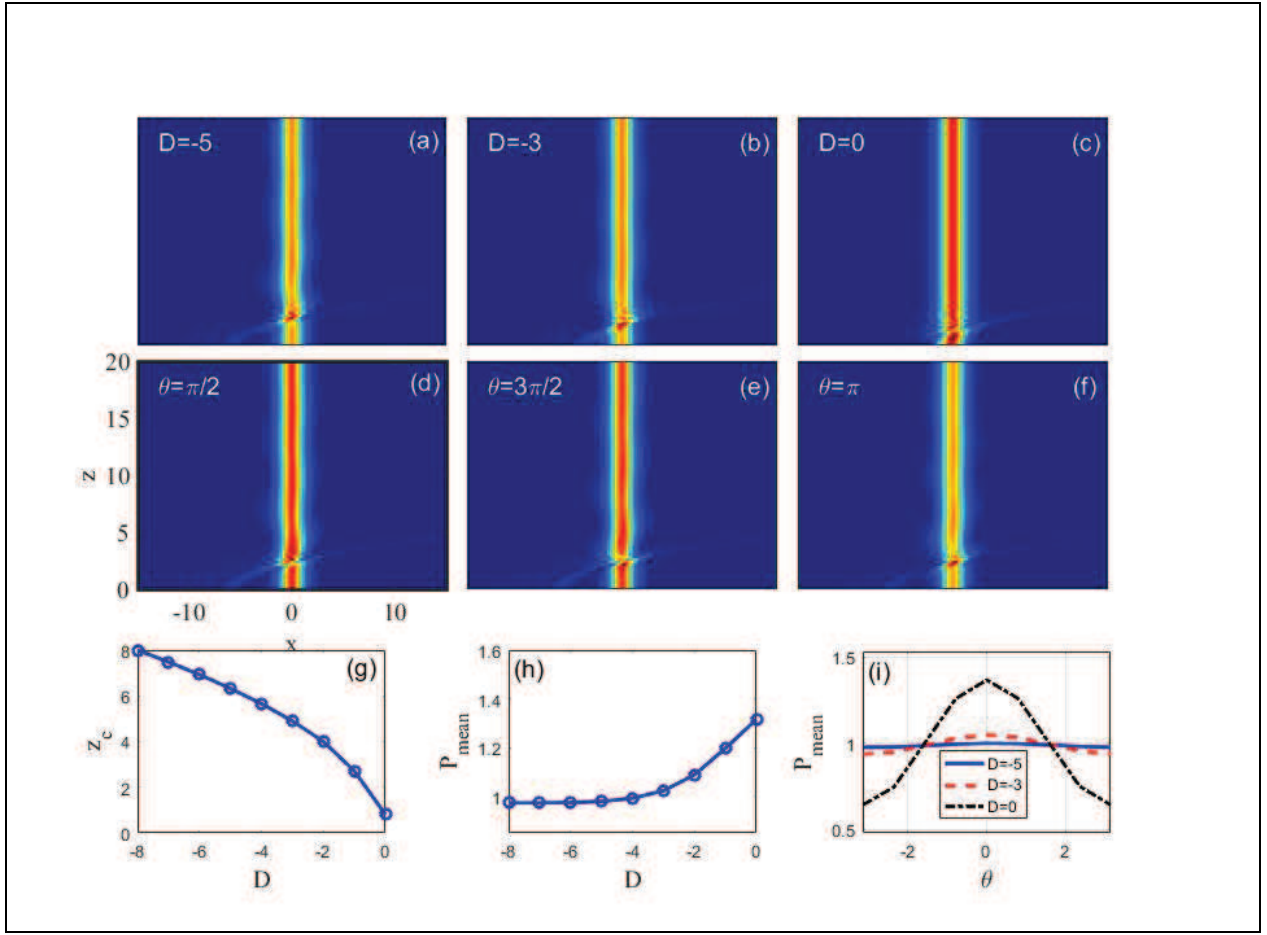}}
\caption{(color online) Airy-soliton interactions with an initial separation of (a) $D=-5$, (b) $D=-3$, and (c) $D=0$ with $\theta=0$ and (d) $\theta=\pi/2$, (e) $\theta=\pi$, and (f) $\theta=3\pi/2$ with $D=-5$. (g) The collision distance $Z_c$ as a function of parameter $D$.  The mean soliton intensity $P_{mean}$ via (h) $D$ and (i) $\theta$, respectively. The other parameters are $q_{A0}=1$, $d=30\mu$m, $V_0=3$, and $W_0=0.3$.}
\label{fig:five}
\end{figure}
\section{Numerical results of Airy-soliton interactions in PT potentials}
\label{section:four}
For Airy-soliton interactions, the influence of the varied beam parameters on the beam propagation is discussed firstly. Fig.~\ref{fig:five} shows that two beams launch at different space separations $D$ or different phase shifts $\theta$ for $q_{A0}=1$, $d=30$, $V_0=3$, and $W_0=0.3$. The propagating Airy beam decelerates to collide with the trailing soliton, so the interaction can be separated to two regimes of interest: the primary collision region between two beams (occurring at approximately $2<z<8$, for $D=-5$ in Fig.~\ref{fig:five}(a)), and a relaxation region accompanied by continuous interaction with the dispersed Airy tail (occurring at $z>8$). These are responsible for the different parameters such as space separation, phase shift, PT potential parameters, and soliton amplitude. During the primary collision, one cannot distinguish two beams which lose their
identities due to interference throughout the collision region ($2<z<8$). However, as the Airy beam further moves towards later positions two beams reform and emerge
having perturbed parameters. In addition, the Airy beam never completely crosses over the soliton, so the Airy-soliton interactions are classified as incomplete collisions~\cite{42}.
From Figs.~\ref{fig:five}(a)-(c), we can say that the collision distance $Z_c$ lessens with $D$ decreasing. $Z_c$ is given by~\cite{42},
\begin{equation}
Z_c=2\sqrt{Z_s-D+Z_{peak}},
\label{eq:six}
\end{equation}
where $Z_s$ is the soliton input position (in our case $Z_s=0$) and $Z_{peak}$ is the offset of the main Airy peak with respect to the space separation $D$. $Z_{peak}$ is numerically calculated for a given truncation and space separation. For example, $Z_{peak}=5$ for $\alpha=0.03$ and $D=-5$, so $Z_c=6.32$. Fig.~\ref{fig:five}(g) verifies the relation of $Z_c$ and $D$ according to Equation~(\ref{eq:six}). The mean soliton power behavior is analyzed for different initial Airy-soliton separations $D$ with in-phase ($\theta=0$) in Fig.~\ref{fig:five}(h).
By establishing soliton power and background power from the maximum and minimum interference values, we can calculate the mean soliton power $P_{mean}$ far from collision. It is obvious that $P_{mean}$ increases with $D$. Moreover, the mean soliton power $P_{mean}$ is charted  for different initial Airy-soliton phases and separations at $q_{A0}=1$, $d=30$, $V_0=3$, and $W_0=0.3$ in Fig.~\ref{fig:five}(i). One can find sinusoidal dependence on the initial Airy phase for all separations, illustrating an energy transfer between the beams during the primary collision. 
\begin{figure}[htb]
\centering
\fbox{\includegraphics[width=\linewidth]{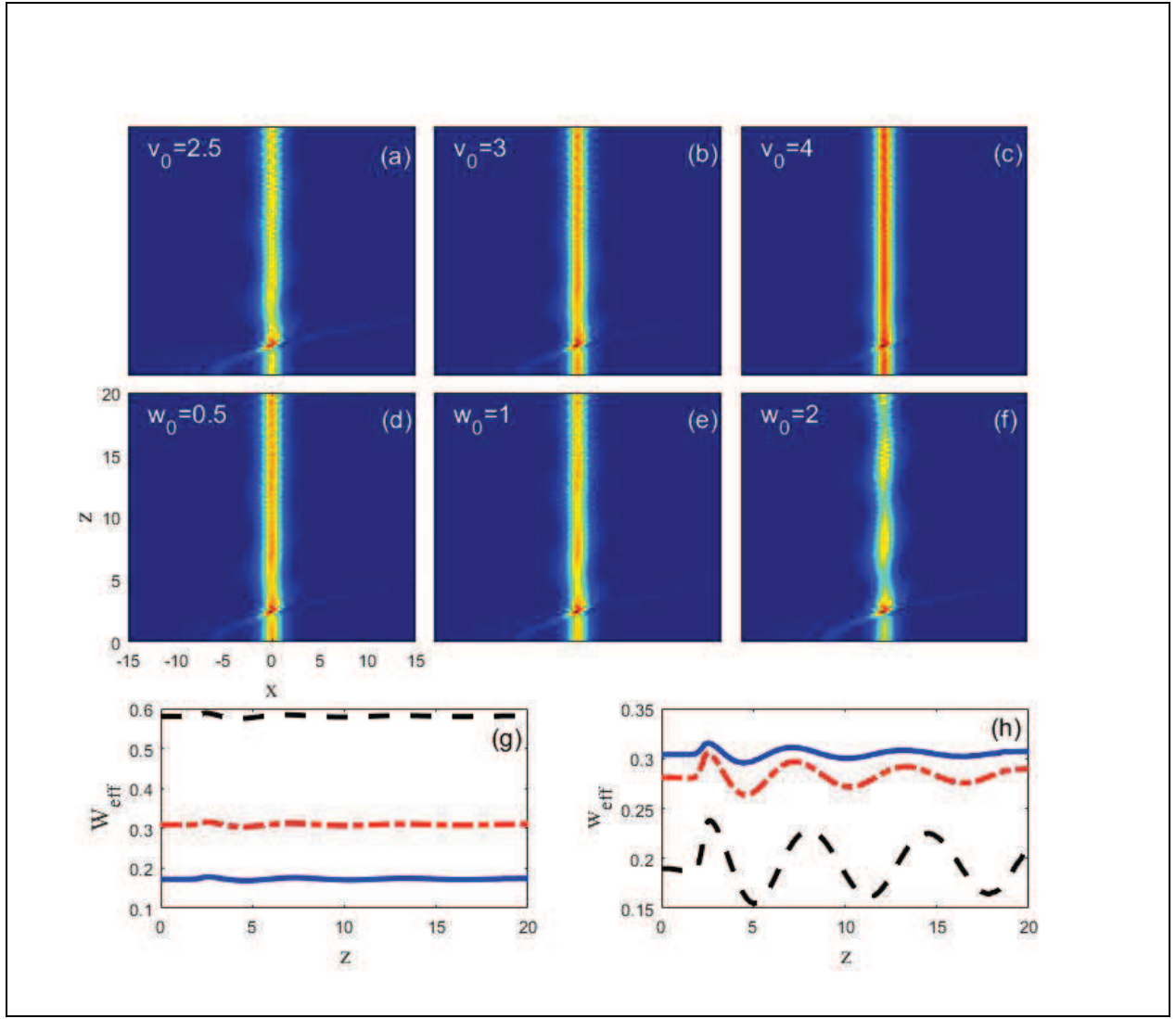}}
\caption{(color online) Intensity plots for the propagation of an Airy beam at (a) $V_{0}=2.5$, (b) $V_{0}=3$, and (c) $V_{0}=4$ with $W_0=0.3$ and (d) $W_{0}=0.5$, (e) $W_{0}=1$, and (f) $W_{0}=2$ for $V_0=3$. (g) The effective beam width $W_{eff}$ via the propagation distance $z$ for $V_0=2.5$ (solid blue line), $V_0=3$ (dash-dotted red line), and $V_0=4$ (dashed black line) at $W_0=0.3$. (h) $W_{eff}$ via $z$ for $W_0=0.5$ (solid blue line), $W_0=1$ (dash-dotted red line), and $W_0=2$ (dashed black line) at $V_0=3$. The other parameters are $q_{A0}=0.5$, $\theta=0$, $D=-5$, and $d=30\mu$m.}
\label{fig:eight}
\end{figure}
Next, we change the modulation amplitudes $V_0$ and $W_0$ of PT potentials to find 
the effect of $V_0$ and $W_0$ on Airy-soliton interactions shown in Fig.~\ref{fig:eight}. From
Figs.~\ref{fig:eight}(a)-(c) and (g), we find the behavior of soliton-like in the relaxation regions for different $V_0$. The beam width of soliton-like $W_{eff}$ increases with the increasing $V_0$. Figs.~\ref{fig:eight}(d)-(f) illustrate the change of the beam propagation with $W_0$. The behaviors are similar to Fig.~\ref{fig:three}(h) in the relaxation regions can be seen in Fig.~\ref{fig:eight}(h) for different $W_0$, which include soliton-like ($W_0=0.5$) and breathers-like ($W_0=2$).

\section{Conclusion}
\label{section:five}
To conclude, we have investigated Airy-soliton interactions in self-defocusing nonlinear media with PT potentials by using the numerical simulations with the split-step Fourier method. If we only consider the propagation of Airy beams in a defocusing nonlinear medium, the propagation dynamics can be  described and divided into three parts due to the existence of PT potentials: One part is soliton-like propagation with small oscillations; The other part stays the same self-accelerating as the input Airy beam; The last part may be associated with reflected waves. The amplitude and period of the oscillations may be affected by input beam amplitude and PT potential parameters. We find that soliton-like and breather-like dynamics can be produced during propagation. For Airy-soliton interactions, the propagating Airy beam can decelerate to collide with the trailing soliton, so the interaction can be separated to two regimes of interest: the primary collision region between the beams and a relaxation region accompanied by continuous interaction with the dispersed Airy tail. In the relaxation region, the behaviors of soliton-like and breather-like dynamics can also be observed. The beam width and mean power are influenced by space separation, phase shift and modulation depth of PT potentials. More interestingly, the collision distance becomes smaller when spatial separation between two beams is smaller, and the mean soliton-like power exhibits sinusoidal dependence on the initial Airy phase.

\section*{Acknowledgments}

 This project was supported by the National Natural Science Foundation of China under (Grant No. 11774068), Natural Science Foundation Guangdong Province of China (Grant No. 2016A030313747), and by the National Key RD Program of China (2017YFA0303800).


\begin{thebibliography}{99}
\bibitem{1}Z. G. Chen, M. Segev, D. N. Christodoulides, ``Optical spatial solitons: historical overview and recent advances," Rep. Prog. Phys. \textbf{75}, 086401 (2012).
\bibitem{2}G. A. Siviloglou and D. N. Christodoulides, ``Accelerating finite energy Airy beams," Opt. Lett. \textbf{32}, 979-981 (2007).
\bibitem{3}G. A. Siviloglou, J. Broky, A. Dogariu, and D. N. Christodoulides, ``Observation of accelerating Airy beams," Phys. Rev. Lett. \textbf{99}, 213901 (2007).
\bibitem{3-1}J. Baumgartl, M. Mazilu, and K. Dholakia, ``Optically mediated particle clearing using Airy wavepackets," Nat. Photonics \textbf{2}, 675-678 (2008).
\bibitem{3-2}R. Schley, I. kaminer, R. Bekenstein, E. Greenfield, Y. Lumer, and M. Segev, ``Loss-proof self-accelerating beams and their use in non-paraxial manipulation of particles¡ä trajectories," Nat. Commu. \textbf{5}, 5189 (2014).
\bibitem{3-3}P. Polynkin, M. Kolesik, J. V. Moloney, G. A. Siviloglou, and D. N. Christodoulides, ``Curved Plasma Channel Generation Using Ultraintense Airy Beams," Science \textbf{324}, 229-232 (2009).
\bibitem{3-4}A. Chong, W. H. Renninger, D. N. Christodoulides, and F. W. Wise, ``Airy-Bessel wave packets as versatile linear light bullets," Nat. Photonics \textbf{4}, 103-106 (2010).
\bibitem{3-5}N. Wiersma, N. Marsal, M. Sciamanna, and D. Wolfersberger, ``All optical interconnects using Airy beams," Opt. Lett. \textbf{39}, 5997-6000 (2014).
\bibitem{3-6}Y. Liang, Y. Hu, D. Song, C. B. Lou, X. Z. Zhang, Z. G. Chen, and J. J. Xu ``Image signal transmission with Airy beams," Opt. Lett. \textbf{40}, 5686-5689 (2015)11.
\bibitem{3-7}S. Jia, J. C. Vaughan, X. Zhuang. ``Isotropic three dimensional super-resolution imaging with a self-bending point spread function," Nat. Photonics \textbf{8}, 302-306 (2014).
\bibitem{3-8}Y. Hu, A. Tehranchi, S. Wabnitz, ``Improved intrapulse Raman scattering control via asymmetric Airy pulses," Phys. Rev. Lett. \textbf{114}, 073901 (2015).
\bibitem{11}I. Kaminer, M. Segev, and D. N. Christodoulides, ``Self-Accelerating Self-Trapped Optical Beams,'' Phys. Rev. Lett. \textbf{106}, 213903 (2011).
\bibitem{12}A. Lotti, D. Faccio, A. Couairon, D. G. Papazoglou, P. Panagiotopoulos, D. Abdollahpour, and S. Tzortzakis, ``Stationary nonlinear Airy beams,'' Phys. Rev. A \textbf{84}, 021807 (2011).
\bibitem{13}I. Kaminer, J. Nemirovsky, and M. Segev, ``Self-accelerating self-trapped nonlinear beams of Maxwell¡¯s equations,'' Opt. Express \textbf{20}, 18827-18835 (2012).
\bibitem{14}P. Zhang, Y. Hu, D. Cannan, A. Salandrino, T. Li, R. Morandotti, X. Zhang, and Z. Chen, ``Generation of linear and nonlinear nonparaxial accelerating beams,'' Opt. Lett. \textbf{37}, 2820-2822 (2012).
\bibitem{15}R. Driben and T. Meier, ``Nonlinear dynamics of Airy-vortex 3D wave packets: emission of vortex light waves,'' Opt. Lett. \textbf{39}, 5539-5542 (2014).
\bibitem{17}S. Jia, J. Lee, J. W. Fleischer, G. A. Siviloglou, and D. N. Christodoulides, ``Diffusion-trapped Airy beams in photorefractive media,'' Phys. Rev. Lett. \textbf{104}, 253904 (2010).
\bibitem{18}R. Bekenstein and M. Segev, ``Self-accelerating optical beams in highly nonlocal nonlinear media,'' Opt. Express \textbf{19}, 23706-23715 (2011).
\bibitem{19}G. Zhou, R. Chen, and G. Ru, ``Propagation of an Airy beam in a strongly nonlocal nonlinear media,'' Laser Phys. Lett. \textbf{11} 105001 (2014).
\bibitem{22}I. Dolev, I. Kaminer, A. Shapira, M. Segev, and A. Arie, ``Experimental Observation of Self-Accelerating Beams in Quadratic Nonlinear Media,'' Phys. Rev. Lett. \textbf{108}, 113903 (2012).
\bibitem{24}I. M. Allayarov and E. N. Tsoy, ``Dynamics of Airy beams in nonlinear media,'' Phys. Rev. A \textbf{90}, 023852 (2014).
\bibitem{25}Y. Fattal, A. Rudnick, and D. M. Marom, ``Soliton shedding from Airy pulses in Kerr media,'' Opt. Express \textbf{19}, 17298-17307 (2011).
\bibitem{26}D. Abdollahpour, S. Suntsov, D. G. Papazoglou, and S. Tzortzakis, ``Spatiotemporal airy light bullets in the linear and nonlinear regimes,'' Phys. Rev. Lett. \textbf{105}, 253901 (2010).
\bibitem{27}P. Panagiotopoulos, D. G. Papazoglou, A. Couairon, and S. Tzortzakis, ``Sharply autofocused ring-Airy beams transforming into non-linear intense light bullets,'' Nat. Commun. \textbf{4}, 2622 (2013).
\bibitem{28}C. Hang and G. Huang, ``Guiding ultraslow weak-light bullets with Airy beams in a coherent atomic system,'' Phys. Rev. A \textbf{89}, 013821 (2014).
\bibitem{28-1}T. Ellenbogen, N. Voloch-Bloch, A. Ganany-Padowicz, and A. Arie, ``Nonlinear generation and manipulation of Airy beams,'' Nat. Photonics \textbf{3}, 395-398 (2009).
\bibitem{28-2}R. P.Chen, C. F. Yin, X. X. Chu, . ``Effect of Kerr nonlinearity on an Airy beam,'' Phys. Rev. A \textbf{82}, 043832 (2010)
\bibitem{28-3}Y. Hu, Z. Sun, D. Bongiovanni, ``Reshaping the trajectory and spectrum of nonlinear Airy beams,'' Opt. Lett. \textbf{37}, 3201-3203 (2012).
\bibitem{28-4}N. K. Efremidis, ``Airy trajectory engineering in dynamic linear index potentials,'' Opt. Lett. \textbf{36}, 3006-3008 (2011).
\bibitem{28-5}N. K. Efremidis, ``Accelerating beam propagation in refractive-index potentials,'' Phys. Rev. A \textbf{89}, 023841 (2014).
\bibitem{28-6}S. Ch\'{a}vez-Cerda, U. Ruiz, V. Arriz\'{o}n, and H. M. Moya-Cessa, ``Generation of Airy solitary-like wave beams by acceleration control in inhomogeneous media,'' Opt. Express \textbf{19}, 16448-16454 (2011).
\bibitem{29}K. G. Makris, I. Kaminer, R. El-Ganainy, N. K. Efremidis, Z. Chen, M. Segev, and D. N. Christodoulides, ``Accelerating diffraction-free beams in photonic lattices,'' Opt. Lett. \textbf{39}, 2129-2132 (2014).
\bibitem{30}Y. Hu, S. Huang, P. Zhang, C. Lou, J. Xu, and Z. Chen, ``Persistence and breakdown of Airy beams driven by an initial nonlinearity,'' Opt. Lett. \textbf{35}, 3952-3954 (2010).
\bibitem{31}Z. Ye, S. Liu, C. Lou, P. Zhang, Y. Hu, D. Song, J. Zhao, and Z. Chen, ``Acceleration control of Airy beams with optically induced refractive-index gradient,'' Opt. Lett. \textbf{36}, 3230-3232 (2011).
\bibitem{32}N. M. Lu\v{c}i\'{c}, B. M. Boki\'{c}, D. \v{Z}. Gruji\'{c}, D. V. Panteli\'{c}, B. M. Jelenkovi\'{c}, A. Piper, D. M. Jovi\'{c}, and D. V. Timotijevi\'{c}, ``Defect-guided Airy beams in optically induced waveguide arrays,'' Phys. Rev. A \textbf{88}, 063815 (2013).
\bibitem{33}A. Piper, D. V. Timotijevi\'{c}, and D. M. Jovi\'{c}, ``Acceleration control of Airy beams with optically induced photonic lattices,'' Phys. Scr. \textbf{T157}, 014023 (2013).
\bibitem{34}F. Diebel, B. M. Boki\'{c}, M. Boguslawski, A. Piper, D. V. Timotijevi\'{c}, D. M. Jovi\'{c}, and C. Denz, ``Control of Airy-beam self-acceleration by photonic lattices,'' Phys. Rev. A \textbf{90}, 033802 (2014).
\bibitem{35}Z. H. Musslimani, K. G. Makris, R. El-Ganainy, and D. N. Christodoulides, ``Optical Solitons in PT Periodic Potentials,'' Phys. Rev. Lett. \textbf{100}, 030402 (2008).
\bibitem{36}H. Wang and J. D. Wang, ``Defect solitons in parity-time periodic potentials,'' Opt. Express \textbf{19}, 4030 (2011).
\bibitem{37}Z. H. Musslimani, K. G. Makris, R. El-Ganainy and D. N. Christodoulides, ``Analytical solutions to a class of nonlinear Schr\"{o}dinger equations with PT-like potentials,'' J. Phys. A \textbf{41}, 244019 (2008).
\bibitem{38}Z. W. Shi, X. J. Jiang, X. Zhu, and H. G. Li, ``Bright spatial solitons in defocusing Kerr media with PT-symmetric potentials,'' Phys. Rev. A \textbf{88}, 053855 (2011).
\bibitem{42}A. Rudnick and D. M. Marom, ``Airy-soliton interactions in Kerr media,'' Opt. Express \textbf{19}, 25570-25582 (2011).
\bibitem{43}W. Cai, M. S.Mills, D. N.Christodoulides, and S.Wen, ``Soliton manipulation using Airy pulses,'' Opt. Commun. \textbf{316}, 127-131 (2014).
\bibitem{44}G. Assanto, A. A.Minzonib, and N. F. Smyth, ``On optical Airy beams in integrable and non-integrable systems,'' Wave Motion \textbf{52}, 183-193 (2015).
\bibitem{45}Z. W. Shi, X. Jing, X. Zhu, Y. Li, and H. G. Li, ``Propagation of an Airy-Gaussian beam in defected photonic lattices,'' Appl. Phys. B \textbf{123}, 159 (2017).
\end{thebibliography}
\end{document}